# Twist-tuned suppression of higher-order modes in single-ring hollow-core photonic crystal fibers

N. N. Edavalath,* M. C. Günendi, R. Beravat, G. K. L. Wong, M. H. Frosz, J.-M. Ménard, P. St.J. Russell

*Max Planck Institute for the Science of Light, Staudtstr. 2, 91052 Erlangen, Germany*
*Corresponding author: nitin.edavalath@mpl.mpg.de*

**Optimum suppression of higher order modes in single-ring hollow-core photonic crystal fibers (SR-PCFs) occurs when the capillary-to-core diameter ratio $d/D$ = 0.68. Here we report that, in SR-PCFs with sub-optimal values of $d/D$, higher-order mode suppression can be recovered by spinning the preform during fiber drawing, thus introducing a continuous helical twist. This geometrically increases the effective axial propagation constant (initially too low) of the $LP_{01}$-like modes of the capillaries surrounding the core, enabling robust single-mode operation. The effect is explored by means of extensive numerical modeling, an analytical model and a series of experiments. Prism-assisted side-coupling is used to investigate the losses and near-field patterns of individual fiber modes in both the straight and twisted cases. More than 12 dB/m improvement in higher order mode suppression is achieved experimentally in a twisted PCF. The measurements also show that the higher order mode profiles change with twist rate, as predicted by numerical simulations. Helical twisting offers an additional tool for achieving effectively endlessly single-mode operation in hollow-core SR-PCFs.**

http://dx.doi.org/

Hollow-core photonic crystal fiber (HC-PCF) [1] has opened up a wide range of new applications for optical fibers, for example in the delivery of high-power laser light, pulse compression [2,3] and enhanced gas-light interactions [4,5]. HC-PCFs are primarily divided into two types based on their guidance mechanism, namely photonic bandgap and anti-resonant reflection (ARR) [6]. A common example of ARR-PCFs is kagomé-PCF [7], which offers relatively low transmission loss over broad spectral ranges in the visible and near-infrared. Stacking a kagomé-PCF preform requires a large number of capillaries, however, rendering the production cumbersome and time consuming. In addition, optical resonances in the many struts and junctions in the extended kagomé-PCF cladding phase-match to the core mode at certain wavelengths, giving rise to loss peaks and degrading the flatness of the transmission spectrum.

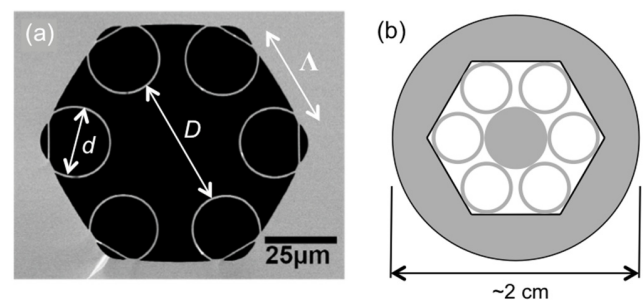

**Fig. 1.** (a) Scanning electron micrograph of a single-ring PCF with core diameter $D$ = 43 μm, capillary diameter $d$ = 23 μm and capillary wall thickness $t$ = 0.7 μm. The inter-capillary distance $\Lambda = (D+d)/2$. (b) Schematic of the initial silica SR-PCF stack, consisting of a jacket tube with a hexagonal inner bore and six capillaries held in place at the vertices by a central glass rod inserted at each end of the stack.

Recently, simpler ARR-PCF structures have been introduced in which the number of cladding unit cells is reduced to a single ring of capillaries surrounding the core (Fig. 1(a)) [8-10]. These capillaries are designed to be anti-resonant elements that block leakage of the $LP_{01}$-like core mode [11,12]. Impressively low losses can be reached with these very simple structures [10].

Although single-ring PCF (SR-PCF) is easier to fabricate than the structurally more complex kagomé-PCF, and does not suffer from a multiplicity of cladding resonances, it generally supports several higher-order modes (HOMs) with losses that are often comparable to that of the $LP_{01}$-like core mode. This makes it difficult to launch a pure fundamental mode, as required in many applications. Several recent studies have tried to address this issue using different approaches.

One such technique is based on phase-matching the $LP_{11}$-like core mode to the $LP_{01}$-like modes of the capillaries, thus providing a leakage channel for HOMs. Optimum HOM suppression turns out to occur at $d/D = 0.68$, where $d$ is the inner diameter of the capillaries and $D$ the core diameter [13]. The fabrication tolerances for reaching this condition are, however, very stringent (often the capillaries end up being too small). It is also important to suppress phase-matching to resonances in the capillary walls, which give rise to bands of high loss in the transmission spectrum. The fundamental (longest wavelength) band occurs at $\lambda_0 = 2h(n_g^2-1)^{0.5}$ where $n_g$ is the refractive index of the glass (~1.46) and $h$ the wall thickness. For example, if it is desired to guide light down to 400 nm, $h$ must be thinner than 188 nm. The requirement for very thin capillary walls makes it even more challenging to reach $d/D = 0.68$. When $d/D$ is sub-optimal, the modal refractive index of the LP$_{11}$-like core mode is higher than the index of the LP$_{01}$-like capillary modes.

Here we report that this difficulty can be overcome by introducing a helical twist, with a precise pitch, during the drawing process. For an off-axis field lobe that is forced by the microstructure to follow a spiral path, twisting has the effect of geometrically increasing its effective path-length (and thus its modal index) along the fiber axis by a factor $(1+\alpha^2\rho^2)^{0.5}$, where $\alpha = 2\pi/L$ is the twist rate, $L$ the helical pitch and $\rho$ is the distance from the fiber axis [14,15]. As a result, the index of the capillary modes is increased by a factor $(1+\alpha^2(d+D)^2/4)^{0.5}$ and that of the LP$_{11}$-like core mode by a factor $(1+\alpha^2\rho_L^2)^{0.5}$, where $\rho_L = \gamma D$ is the radial position of the lobes and $\gamma$ is a numerical factor (see below).

In addition, because its individual field lobes are strongly coupled to each other, the LP$_{11}$-like core mode will transform into a helical Bloch mode in the twisted fiber, with azimuthal harmonics of orbital angular momentum (OAM) order $\ell^{(m)} = \ell_0 + mN$ where $\ell_0$ is the order of the principal (strongest) harmonic, $m$ the harmonic order, and $N$ the number of field lobes [16]. The $m$-th harmonic will have an axial refractive index greater by an amount $\ell^{(m)}\alpha\lambda/2\pi$ than in the straight fiber (this does not occur in the ring of capillaries because they are only very weakly coupled, if at all).

Combining all these effects yields a condition at which the indices of the higher order core mode and capillary mode match:

$$n_{01}\sqrt{1+\alpha^2(d+D)^2/4} = \ell^{(m)}\alpha\lambda/(2\pi) + n_{11}\sqrt{1+\alpha^2\gamma^2 D^2} \quad (1)$$

where

$$n_{01} = \sqrt{1-\left(u_{01}\lambda/(\pi f_{01}d)\right)^2} \text{ and } n_{11} = \sqrt{1-\left(u_{11}\lambda/(\pi f_{11}D)\right)^2} \quad (2)$$

are the modal indices of the LP$_{01}$ and LP$_{11}$ modes in a straight capillary, $u_{kl}$ is the $l$-th zero of a Bessel function of order $k$ and the correction factors $f_{11} = 1.077$ and $f_{01} = 0.991$ [17]. Recognizing that the second terms under the square-roots are all much less than unity and taking terms up to first order, Eq. (1) can be re-cast in the form:

$$q^2 - q\left(\frac{2f_{01}\ell^{(m)}}{u_{01}\left((\xi+1)^2-4\gamma^2\right)}\right) - \left(\frac{\xi_0^2-\xi^2}{\xi^2\xi_0^2\left((\xi+1)^2-4\gamma^2\right)}\right) = 0 \quad (3)$$

where

$$q = \pi\alpha f_{01}D^2/(2\lambda u_{01}) \quad (4)$$

and $\xi_0 = u_{01}f_{11}/(u_{11}f_{01})$ is the optimal value of $\xi = d/D$ at zero twist rate.

Fig. 2(a) plots the positive-valued solution of Eq. (3), together with the results of finite element modeling. Excellent agreement is obtained for $\gamma = 0.175$ and $\ell^{(m)} = -1$, i.e., for the lowest index Bloch harmonic that has an appreciable amplitude. The optimal twist rate for the experimental parameters ($\lambda = 1064$ nm and $D = 43$ μm) is shown on the right-hand axis.

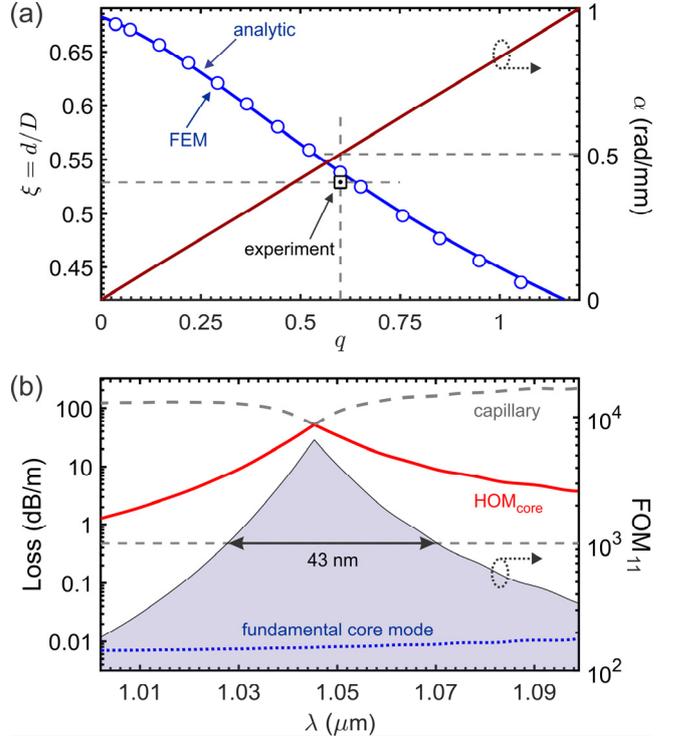

**Fig. 2.** (a) The value of $d/D$ that provides optimal HOM suppression, plotted against the parameter $q$ (see text). The marked point correspond to the experimental value of $d/D = 0.53$, for which the experimental twist rate (505 rad/m) is quite close to the optimal value of 563 rad/m predicted by Eq. (3). (b) Wavelength dependence of the loss (left axis) of the LP$_{01}$-like core mode (dotted), the LP$_{11}$-like core mode (solid red) and the capillary modes (dashed). The corresponding figure of merit (FOM) is shown on the right axis. The bandwidth is ~43 nm for FOM$_{11}$ > 1000.

Armed with this analytical model as a guide, we carried out a series of experiments and numerical simulations, comparing the properties of the modes in twisted and untwisted versions of the same single-ring HC-PCF structure.

The fibers were fabricated using a modified version of the two-step stack-and-draw technique. The structure was first stacked by placing capillaries at the vertices of a jacket tube with a hexagonal inner profile (Fig. 1(b)). The capillaries were held in place by two central support rods, one at each end of the stack. This structure was first drawn down to a intermediate cane of outer diameter ~2 mm. The intermediate cane was then spun around its axis while being drawn down to fiber [16]. The twist rate is determined by the drawing and rotation speeds, and long lengths (limited only by size of the cane) of twisted single-ring HC-PCF could be drawn with minimal structural distortion. Two samples were used in the experiments: one was untwisted and the other had a twist-rate of 505 rad/m. Both fibers had closely similar structural parameters, namely $D = 43$ μm, $h = 0.7$ μm, and $d = 23$ μm, yielding $d/D = 0.53$, which is sub-optimal for HOM suppression.

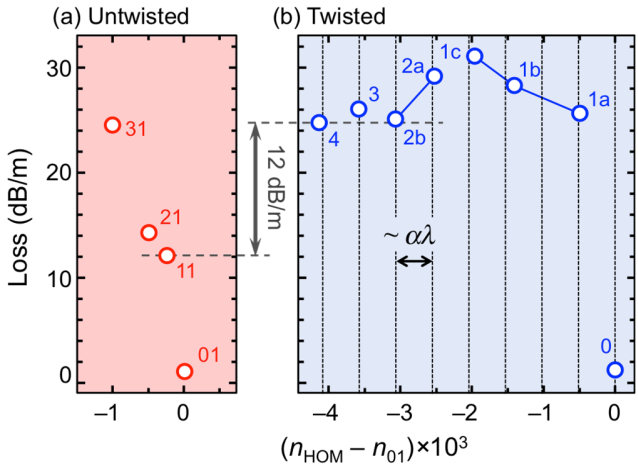

**Fig. 3.** Experimentally measured (using prism coupling) higher-order mode loss plotted versus modal refractive index relative to the fundamental mode for an ARR-PCF with $d/D$ = 0.53. (a) Modal loss and indices for the untwisted case. The labels next to the data-points for the red curve indicate the azimuthal $k$ and radial $l$ orders of the $LP_{kl}$-like modes. (b) Twist rate 505 rad/m. Modes with similar near-field distributions (see Fig. 4) are grouped together. The spacing between successive modes is approximately a multiple of $\alpha\lambda$ (see text). The loss of HOMs is increased by at least 12 dB/m in the twisted fiber.

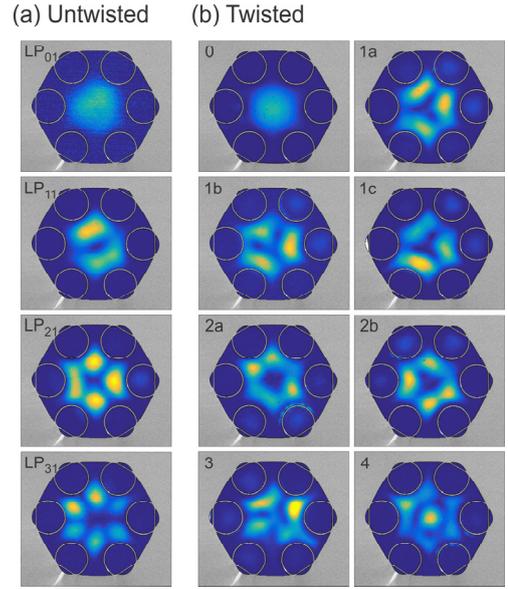

**Fig. 4.** Measured optical near-field distributions of modes excited by prism-assisted side-coupling at 1064 nm in 60 cm lengths of (a) untwisted and (b) twisted ($\alpha$ = 505 rad/m) SR-PCF. The mode profiles are superimposed on a scanning electron micrograph of the fiber structure. The corresponding modal losses and refractive indices are plotted in Fig. 3.

The fiber modes were characterized using prism-assisted side-coupling, which allows light to be fed precisely into a selected individual fiber mode, provided it is leaky enough [18]. The effective modal refractive indices, losses and associated modal patterns of HOMs can be studied in detail using this approach. The experiment involves placing a wedge prism (face slanted at 1.1°) on the side of the SR-PCF with a layer of index-matching fluid in between. Light from a 1064 nm diode laser is coupled into a commercially available single-mode fiber (SMF), the output of which is collimated to a full-width-half-maximum beam-width of 2 mm. This beam is then arranged to be incident on the slanted face of the wedge prism, at an angle $\psi$ to its face normal, and is focused on to a line parallel to the fiber axis using a cylindrical lens (focal length 30 mm). The single-mode fiber and the lens are mounted together on a rotation stage so as to allow $\psi$ to be precisely varied.

As $\psi$ is increased, successive higher order modes are excited when the wavevector component along the fiber axis matches the wavevector of a mode. The near-field intensity profiles of the modes can then be imaged at the fiber end using a CCD camera. Unlike end-fire coupling, where precise matching of the modal profile is essential if excitation of unwanted modes is to be avoided, prism-assisted side-coupling allows pure higher-order modes to be selectively excited without prior knowledge of their field profile. Once a mode is optimally excited, its modal refractive index can be calculated [18] from the measured beam angle, taking careful account of refractive indices at 1064 nm (1.4496 for the silica glass prism at 1064 nm and 1.0003 for air). The procedure is then repeated for different fiber lengths, allowing the loss of each mode to be evaluated with good accuracy.

Fig. 3 plots the measured propagation losses and modal refractive indices of the modes in twisted and untwisted SR-PCF, and the corresponding near-field mode profiles are shown in Fig. 4. In the untwisted case it is fairly straightforward to identify the modes and label them using the $LP_{kl}$ notation, where $k$ and $l$ are the azimuthal and radial orders respectively (Figs. 3(a) and 4(a)).

In the twisted case, however, it is more difficult to identify the HOMs using the LP notation, because they are strongly affected by the twist. It is nevertheless possible to arrange them in groups with similar-looking field profiles, as is done in Fig. 4(b). Group 1 (containing 1a, 1b and 1c) modes have three-fold symmetry, whereas Group 2 (2a and 2b) modes have six-fold symmetry. Mode 3 is somewhat difficult to identify, but mode 4 clearly has some similarity with an $LP_{32}$ mode.

To understand this behavior it is necessary first to realize that the both axial and azimuthal phase-matching must be satisfied for prism coupling to work. The additional azimuthal component of momentum comes from the slanted capillaries, which act as a long-period grating (Fig. 5). The resulting azimuthal component of refractive index can be written in the form:

$$n_{az}^m = \frac{m\lambda}{\Lambda}\cos\theta - n_{\text{prism}}\sin\phi \simeq \frac{m\lambda}{\Lambda} - n_{\text{prism}}\sin\phi \quad (5)$$

where $m$ is the diffracted order, $n_{\text{prism}}$ the index set by the prism, $\theta \simeq \alpha\Lambda$ the slant angle of the capillaries and $\phi$ is a possible angular deviation of the incident rays from the fiber axis, caused by prism misalignment or the angular spread of the incident beam.

In order to excite an OAM mode of order $\ell$, $m$ and $\phi$ must satisfy the following equation:

$$\ell = N\Lambda'\left(\frac{n_{\text{prism}}\sin\phi}{\lambda} + \frac{m}{\Lambda}\right) \quad (6)$$

where $\Lambda'$ is the distance between adjacent field lobes of a core mode with *N*-fold rotational symmetry, i.e., $N\Lambda' \approx 2\pi\gamma D$. The axial refractive index component of the mode is given by

$$n_z \simeq n_{\text{prism}} + m\alpha\lambda \quad (7)$$

where the approximation $\cos\phi \simeq 1$ has been used. This equation shows that the modal index measured by the prism will be smaller than its actual value by $m\alpha\lambda$, where *m* increases by approximately 1 for modes with successively higher OAM order (note that negative values of *m*, which in our definition imply higher values of axial index, are unlikely to result in excitation of a mode because there is less field penetration outside the core, making prism coupling much less efficient). The predictions of Eq. (7) are confirmed by the measurements in Fig. 3(b), where the index spacing between successive modes corresponds approximately to multiples of $\alpha\lambda = 5.3\times10^{-4}$. Between modes 1a and 1b the index measured by the prism falls by roughly twice this amount, which we tentatively attribute to uncertainties in the angle *ϕ*.

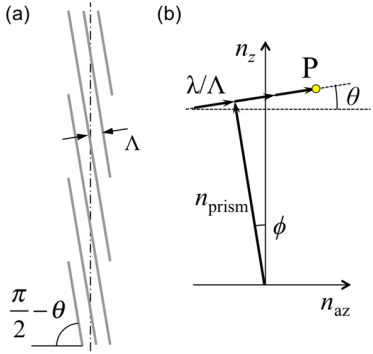

**Fig. 5.** (a) Top view of the capillaries, spaced by $\Lambda = (D+d)/2$ and slanted at an angle $\phi \approx \alpha\Lambda$ to the fiber axis. (b) Refractive index diagram of the grating. At point P the azimuthal component of refractive index is $n_{\text{az}}(P) = (m\lambda/\Lambda)\cos\theta - n_{\text{prism}}\sin\phi$ with *m* = 2, where $n_{\text{prism}}$ is the index set by the prism and *ϕ* is a small angular misalignment (see text). The axial component of refractive index at P is $n_z \simeq n_{\text{prism}} + m\alpha\lambda$.

Although all of the HOMs experience higher loss in the twisted fiber, for pure single-mode operation it is most important that the $LP_{11}$-like mode is strongly suppressed, since it is the one most easily excited by end-fire coupling. Comparing the twisted and untwisted cases (Fig. 3), it is clear that the $LP_{11}$ mode is indeed most strongly suppressed by twisting. The overall HOM loss increase, achieved experimentally, is better than 12 dB/m. In Fig. 2(b) the wavelength dependence of the figure-of-merit for suppression of the $LP_{11}$-like mode, $FOM_{11} = \alpha_{11}/\alpha_{01} - 1$ [13], is calculated numerically, showing that the bandwidth within which $FOM_{11} > 1000$ is 43 nm.

It is interesting that the $LP_{11}$-like mode in the untwisted fiber is double-lobed whereas the first three OAM modes in the twisted fiber have a three-lobed pattern (Fig 4(b)). The evolution of the modal Poynting vector distribution of the first higher order mode with increasing twist rate is explored numerically in Fig. 6. It evolves from a double-lobed pattern, through a doughnut-shape at 150 rad/m, to a triple-lobed mode at 505 rad/m. The fundamental $LP_{01}$-like mode remains unaffected by the twist.

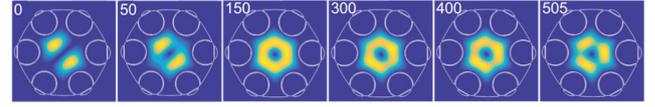

**Fig. 6.** Numerically modeled Poynting vector distributions showing how the double-lobed $LP_{11}$-like mode of the untwisted fiber evolves into a triple-lobed pattern in the twisted fiber as twist rate increases (the values correspond to rad/m).

The use of a preform tube with a hexagonal inner bore facilitates construction of a single-ring preform stack with six cladding capillaries. Spinning the preform cane during fiber drawing permits fabrication of long lengths of azimuthally periodic single-ring hollow photonic crystal fiber that is helically twisted around its axis. The twist-induced geometrical increase in path-length of the capillary modes makes it possible to phase-match the unwanted $LP_{11}$-like core mode to the capillary modes in structures with sub-optimal values of $d/D < 0.68$, causing high $LP_{11}$ mode loss and rendering the fiber effectively single-mode. In the fiber investigated, the figure of merit for suppression of higher order modes is higher than 1000 over a bandwidth of 43 nm at 1064 nm wavelength.